\documentclass[aip, reprint, noshowpacs, superscriptaddress]{revtex4-1}
\usepackage[utf8]{inputenc}
\usepackage[english]{babel}
\usepackage{physics}
\usepackage{array}
\usepackage{graphicx}
\usepackage{times}
\usepackage{color}
\definecolor{linkcolor}{rgb}{0,0,0.6}	
\usepackage[colorlinks=true, pdfstartview=FitV, linkcolor= linkcolor, citecolor= linkcolor, urlcolor= linkcolor, hyperindex=true, hyperfigures=false]{hyperref}			
\usepackage[normalem]{ ulem }
\usepackage{soul}  

\addto\captionsenglish{}

\newcommand*{\rightvect}{\vec}
\newcommand*{\leftvect}[1]{
    \reflectbox{\ensuremath{\vec{\reflectbox{\ensuremath{#1\;}}}}}\hspace{-0.25em}
}
\newcommand*{\srightvect}[1]{{\scriptsize \rightvect{#1}}}
\newcommand*{\sleftvect}[1]{{\scriptsize \leftvect{#1}}}

\newcommand*{\e}{\text{e}}
\newcommand*{\ii}{\text{i}}
\newcommand*{\I}{\text{I}}
\newcommand*{\Traj}{[\rightvect{\Sigma}]}
\newcommand*{\rTraj}{[\leftvect{\Sigma}]}

\newcommand*{\Dt}{\Delta t}
\newcommand*{\kT}{k_{\text{B}} T}
\newcommand*{\ds}{\Delta_{\text{i}} s}
\newcommand*{\gm}{g_\text{m}}

\newcommand*{\rqm}{\rho_{\text{qm}}}

\let\supplsection\subsection
\makeatletter

\renewcommand{\section}{\@startsection {section}{1}{0pt}%
    {20pt}{0.01pt}%
    {\sf\bfseries\raggedright}%
}
\renewcommand{\subsection}{\@startsection {subsection}{2}{0pt}%
    {-0pt}{-0.5em}%
    {\bfseries}%
}
\renewcommand{\subsubsection}{\@startsection {subsubsection}{2}{0pt}%
    {5pt}{-0.5em}%
    {\itshape\bfseries}%
}
\makeatother

\begin{document}
\begin{abstract}
    According to the Second Law of thermodynamics, the evolution of physical systems has a preferred direction,
    that is characterized by some positive entropy production. Here we propose a direct way to measure the
    stochastic entropy produced while driving a quantum open system out of thermal equilibrium. 
    The driving work is provided by a quantum battery, the system and the battery forming an autonomous machine. We show that the battery's energy fluctuations
    equal work fluctuations and check Jarzynski's equality. Since these energy fluctuations are measurable, the battery behaves as an embedded quantum work meter and the machine verifies a generalized fluctuation theorem involving the information encoded in the battery. Our proposal can be implemented
    with state-of-the-art opto-mechanical systems. It paves the way towards the experimental demonstration
    of fluctuation theorems in quantum open systems.  
\end{abstract}

\title{An autonomous quantum machine to measure the thermodynamic arrow of time}
\author{Juliette Monsel}
\email{juliette.monsel@neel.cnrs.fr}
\affiliation{Univ. Grenoble Alpes, CNRS, Grenoble INP, Institut N\'eel, 38000 Grenoble, France}
\author{Cyril Elouard}
\affiliation{Univ. Grenoble Alpes, CNRS, Grenoble INP, Institut N\'eel, 38000 Grenoble, France}
\affiliation{Department of Physics and Astronomy, University of Rochester, Rochester, New York 14627, USA}
\author{Alexia Auff\`eves}
\email{alexia.auffeves@neel.cnrs.fr}
\affiliation{Univ. Grenoble Alpes, CNRS, Grenoble INP, Institut N\'eel, 38000 Grenoble, France}
\date{\today}

\maketitle

\noindent Irreversibility is a fundamental feature of our physical world. The degree of irreversibility of thermodynamic transformations is measured by the entropy production, which is always positive according to the Second Law. At the microscopic level, stochastic thermodynamics \cite{sekimoto, thermo_stoc_classique} has extended this concept to characterize the evolution of small systems coupled to reservoirs and driven out of equilibrium. Such systems follow stochastic trajectories $\rightvect{\Sigma}$ and the stochastic entropy production $\ds\Traj$ obeys the integral fluctuation
theorem (IFT) $\ev{ \exp(-\ds\Traj)}_{\rightvect{\Sigma}} = 1 $ where $\ev{\cdot}_{\rightvect{\Sigma}}$ denotes the average over all trajectories 
$\rightvect{\Sigma}$. Jarzynski's equality (JE) \cite{jarzynski} is a paradigmatic example of such IFT, that constrains the fluctuations of the entropy produced while driving some initially thermalized system out of equilibrium. Experimental demonstrations of JE especially require the ability to measure the stochastic work $W\Traj$ exchanged with the external entity driving the system. In the classical regime, $W\Traj$ can be completely reconstructed from the monitoring of the system's trajectory, allowing for successful experimental demonstrations \cite{jarz_elec, jarz_osc_macro,jarzynski_classique_2018}. 

Defining and measuring the entropy production in the quantum regime is of fundamental interest in the perspective of optimizing the performances of quantum heat engines and the energetic cost of quantum information technologies \cite{Mancini_arxiv, Mancini_npjQI, Santos2017a,Francica2017}. However, measuring a quantum fluctuation theorem can be problematic in the genuinely quantum situation of a coherently driven quantum system, because of the fundamental and practical issues to define and measure quantum work \cite{TH2016, BaumerChapter, engel_jarzynski_2007,RMPCampisi}. So far the quantum JE has thus been extended and experimentally verified in {\it closed} quantum systems, i.e. systems that are driven but otherwise isolated. In this case work corresponds to the change in the system's internal energy, accessible by a two-points measurement protocol \cite{TH2016} or the measurement of its characteristic function \cite{Mazzola,Dorner,DeChiaraChapter}. Experimental demonstrations have been realized, e.g. with trapped ions \cite{an_experimental_2015,jarzynski_quantique_2018}, ensemble of cold atoms \cite{cerisola_using_2017}, and spins in Nuclear Magnetic Resonance (NMR) \cite{Serra2014} where the thermodynamic arrow of time was successfully measured \cite{Serra2015}. 

On the other hand, realistic strategies must still be developed to measure the fluctuations of entropy production for quantum {\it open} systems, i.e. that can be simultaneously driven, and coupled to reservoirs. Since work is usually assumed to be provided by a classical entity, most theoretical proposals so far have relied on the measurement of {\it heat} fluctuations, i.e. small energy changes of the reservoir. Experimentally, this requires to engineer this reservoir and to develop high efficiency detection schemes, which is very challenging \cite{pekola, Horowitz12, elouard_probing_2017}. Experimental demonstrations have remained elusive.

In this article, we propose a new and experimentally feasible strategy to measure the thermodynamic arrow of time for a quantum open system in Jarzynski's protocol, that is based on the direct measurement of {\it work} fluctuations. We investigate a so-called hybrid opto-mechanical system\cite{Treutlein}, that consists in a two-level system (further called a qubit) strongly coupled to a mechanical oscillator (MO) on the one hand, and to a thermal bath on the other hand. Studying single quantum trajectories of the hybrid system, we show that the MO and the qubit remain in a product state all along their joint evolution, allowing to unambiguously define their stochastic energies. We evidence that the mechanical energy fluctuations can be identified with the stochastic work received by the qubit and satisfy JE. Therefore the MO plays the role of a quantum battery, the ensemble of the qubit and the battery forming an autonomous machine \cite{frenzel_quasi-autonomous_2016,tonner_autonomous_2005,holmes_coherent_2018}. Originally, the battery behaves as an embedded quantum work meter, encoding information on the stochastic work exchanges. We show that the evolution of the complete machine is characterized by a generalized IFT, that quantitatively involves the amount of extracted information. This situation gives rise to so-called absolute irreversibility, in agreement with recent theoretical predictions and experimental results \cite{murashita_nonequilibrium_2014,funo2015,Ueda_PRA,Nakamura_arXiv}. Our proposal is robust against finite measurement precision \cite{lahaye_approaching_2004, schliesser_resolved-sideband_2009} and can be probed with state-of-the-art experimental devices. 

The paper is divided as follows. Firstly, we introduce hybrid opto-mechanical devices as autonomous machines, and build the framework to model their evolution on average and at the single trajectory level. Focusing on Jarzynski's protocol, we define stochastic heat, work and entropy production and study the regime of validity and robustness of JE as a function of the parameters of the problem and experimental imperfections. Finally, we derive and simulate an IFT for the complete machine, evidencing the presence of absolute irreversibility. Our results demonstrate that work fluctuations can be measured directly, by monitoring the energetic fluctuations of the quantum battery. They represent an important step towards the experimental demonstration of quantum fluctuation theorem in a quantum open system.  

\section*{Results} 
\subsection*{Hybrid opto-mechanical systems as autonomous machines.} 
\begin{figure}[htb!]
    \includegraphics[scale=1]{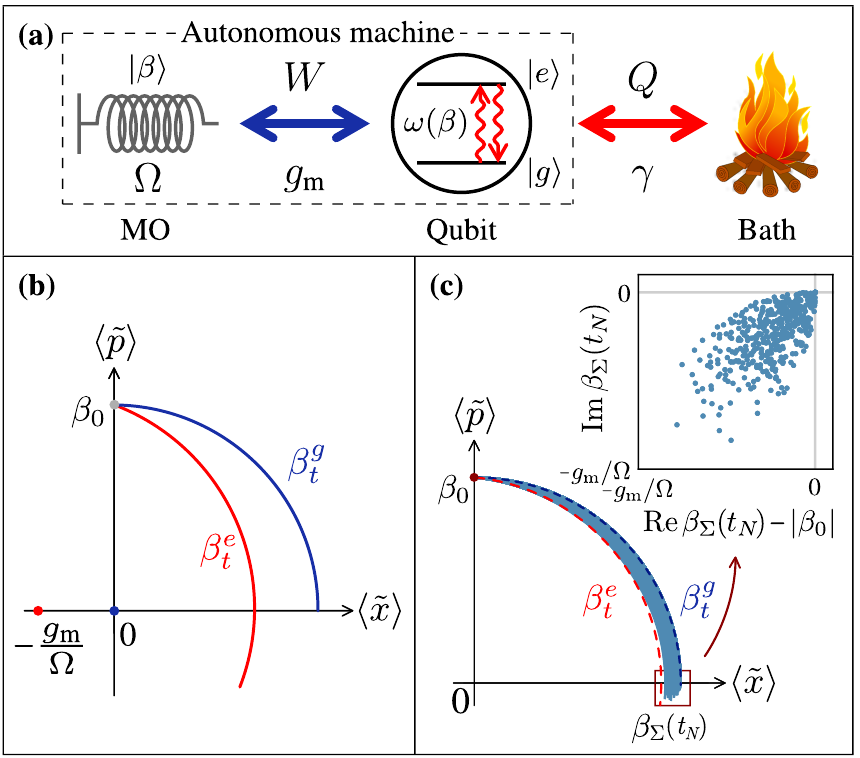}
    \caption{\label{fig1}
        \textbf{(a)} Situation under study: a qubit exchanging work $W$ with a mechanical resonator and heat $Q$ with a thermal bath at temperature $T$. The ensemble of the qubit and mechanics constitutes an autonomous machine. \textbf{(b)} Evolution of the complex mechanical amplitude $\beta$ if the qubit is in the $\ket{e}$ (resp. $\ket{g}$) and the MO is initially prepared in the state $\big|\ii|\beta_0|\big\rangle$. The mechanics can be used as a meter to detect the qubit state if $\gm /\Omega \gg 1$ (ultra-strong coupling regime). The mechanical fluctuations induced by the qubit state are small w.r.t. the free evolution if $|\beta_0| \gg \gm/\Omega$ (semi-classical regime). These two regimes are compatible (See text) \textbf{(c)} Stochastic mechanical trajectories $\rightvect{\beta}[\rightvect{\epsilon}\,]$ in the phase space defined by $(\tilde{x},\tilde{p})$ (See text). The MO is initially prepared in the coherent state $\big|\ii|\beta_0|\big\rangle$, and the qubit state is drawn from thermal equilibrium. Inset: Distribution of final states $\ket{\beta_\Sigma(t_N)}$ within an area of typical width $\gm/\Omega$. Parameters: $T = 80$ K, $\hbar\omega_0 = 1.2\kT$, $\Omega/2\pi = 100$ kHz, $\gamma/\Omega = 5$, $\gm/\Omega = 100$, $|\beta_0| = 1000$.}
\end{figure}

A hybrid opto-mechanical system consists in a qubit of ground (resp. excited) state $\ket{g}$ (resp. $\ket{e}$) and transition frequency $\omega_0$, parametrically coupled to a mechanical oscillator of frequency $\Omega \ll \omega_0$ (See Fig.~\ref{fig1}a). Recently, physical implementations of such hybrid systems have been realized on various platforms, e.g. superconducting qubits embedded in oscillating membranes \cite{hybrid_circuit}, nanowires coupled to diamond nitrogen vacancies \cite{NV_defect}, or to semiconductor quantum dots \cite{trompette}. The complete Hamiltonian of the hybrid system reads $H_{\text{qm}} = H_{\text{q}} + H_{\text{m}} + V_{\text{qm}}$\cite{Treutlein}, where $H_{\text{q}} = \hbar\omega_0 \dyad{e}{e} \otimes  \mathbf{1}_{\text{m}}$ and $H_{\text{m}} =  \mathbf{1}_{\text{q}} \otimes \hbar\Omega b^\dagger b$ are the qubit and MO free Hamiltonians respectively. We have introduced the phonon annihilation operator $b$, and $ \mathbf{1}_{\text{m}}$ (resp. $ \mathbf{1}_{\text{q}}$) the identity on the MO (resp. qubit) Hilbert space. The coupling Hamiltonian is $V_{\text{qm}} = \hbar \gm\dyad{e}{e} \otimes (b + b^\dagger)$, where $\gm$ is the qubit-mechanical coupling strength.  Of special interest for the present paper, the so-called ultra-strong coupling regime is defined as $\gm \geq \Omega$, with $\omega_0 \gg \gm$. It was recently demonstrated experimentally\cite{trompette}.

The Hamiltonian of the hybrid system can be fruitfully rewritten $H_{\text{qm}} = \dyad{e} \otimes H_\text{m}^{e} +  \dyad{g} \otimes H_\text{m}^{g}$ with $H_\text{m}^{g} = \hbar \Omega b^\dagger b$ and $H_\text{m}^{e} =   \hbar \Omega B^\dagger B + \hbar (\omega_0 - \gm^2/\Omega)\mathbf{1}_{\text{m}}$, with $B = b + (\gm/\Omega)\mathbf{1}_{\text{m}}$. It appears that the qubit bare energy states $\epsilon=e,g$ are stable under the dynamics and perfectly determine the evolution of the MO ruled by the Hamiltonian $H_\text{m}^{\epsilon}$. Interestingly, $H_\text{m}^{\epsilon}$ preserves the statistics of coherent mechanical states, defined as $\ket{\beta} = e^{\beta^* b - \beta b^\dagger}\ket{0}$, where $\ket{0}$ is the zero-phonon state and $\beta$ the complex amplitude of the field. Consequently, if the hybrid system is initially prepared in a product state $\ket{\epsilon,\beta_0}$, it remains in a similar product state $\ket{\epsilon, \beta^\epsilon_t}$ at any time, with $\ket{ \beta^\epsilon_t} = \exp(-i H_\text{m}^{\epsilon}t/\hbar) \ket{\beta_0}$. The two possible mechanical evolutions are pictured in Fig.~\ref{fig1}b between time $t_0=0$ and $t = \Omega/2\pi$, in the phase space defined by the mean quadratures of the MO $\langle \tilde{x} \rangle = \langle b+b^\dagger \rangle$ and $\langle \tilde{p} \rangle = -i \langle b-b^\dagger \rangle$. If the qubit is initially prepared in the state $\ket{e}$ (resp. $\ket{g}$), the mechanical evolution is a rotation around around the displaced origin $(-\gm/\Omega,0)$ (resp. the origin $(0,0)$). Such displacement is caused by the force the qubit exerts on the MO, that is similar to the optical radiation pressure in cavity opto-mechanics. Defining $\delta \beta_t = \beta^e_t - \beta^g_t$, it appears that the distance between the two final mechanical states $|\delta \beta_t|$ scales like $\gm/\Omega$. In the ultra-strong coupling regime, this distance is large such that mechanical states are distinguishable, and can be used as quantum meters to detect the qubit state. 

Since the hybrid system remains in a pure product state at all times, its mean energy defined as ${\cal E}_\text{qm}(\epsilon,\beta^\epsilon_t) = \bra{\epsilon,\beta^\epsilon_t} H_\text{qm} \ket{\epsilon,\beta^\epsilon_t}$ naturally splits into two distinct components respectively quantifying the qubit and the mechanical energies:
\begin{align}
{\cal E}_\text{q}(\epsilon,\beta^\epsilon_t) & = \hbar \omega(\beta^\epsilon_t) \delta_{\epsilon,e} \label{Eq} \\
{\cal E}_\text{m}(\beta^\epsilon_t) & = \hbar \Omega |\beta^\epsilon_t|^2, \label{Em}
\end{align}
where $\delta_{\epsilon,e}$ is the Kronecker delta and $\omega(\beta)$ is the effective transition frequency of the qubit defined as:
\begin{equation} \label{omega_eff}
\omega(\beta^\epsilon_t)  = \omega_0 + 2 \gm \Re(\beta^\epsilon_t).
\end{equation}

The frequency modulation described by Eq.~\eqref{omega_eff} manifests the back-action of the mechanics on the qubit. Note that the case $\gm/\Omega \ll |\beta_0|$ corresponds to $|\delta \beta_t| \ll |\beta^g_t|$:  Then the frequency modulation is independent of the qubit state and follows $\omega(\beta^\epsilon_t) \sim \omega(\beta_0 e^{-i\Omega t})$, even in the ultra-strong coupling regime. In what follows, we will be especially interested in the regime where $1\ll \gm/\Omega \ll |\beta_0|$, where the mechanical evolution depends on the qubit state, while the qubit transition frequency is independent of it.\\

We now take into account the coupling of the qubit to a bath prepared at thermal equilibrium. The bath of temperature $T$ consists of a spectrally broad collection of electromagnetic modes of frequencies $\omega'$, each mode containing a mean number of photons $\bar{n}_{\omega'} = \left(\exp(\hbar\omega'/\kT )- 1\right)^{-1}$. The bath induces transitions between the states $\ket{e}$ and $\ket{g}$, and is characterized by a typical correlation time $\tau_c$ giving rise to a bare qubit spontaneous emission rate $\gamma$. 

The hybrid system is initially prepared in the product state $\rho_\text{qm}(0) = \rho_\text{q}(0)\otimes \dyad{\beta_0}$. $\rho_\text{q}(0)$ is the qubit state, taken diagonal in the $\{ e,g \}$ basis. $\dyad{\beta_0}$ is the mechanical state, that is chosen pure and coherent. In the rest of the paper, we shall study transformations taking place on typical time scales $t \sim \Omega^{-1}$, such that the mechanical relaxation is neglected. From the properties of the interaction with the bath and the total hybrid system's Hamiltonian $H_\text{qm}$, it clearly appears that the qubit does not develop any coherence in its bare energy basis. We show in the Supplementary \cite{suppl} that as long as $|\beta_0 | \gg \gm t$, the MO imposes a well defined modulation of the qubit frequency $\omega(\beta_0(t))$ with $\beta_0(t) = \beta_0 e^{-i\Omega t}$. This defines the semi-classical regime, where the hybrid system evolution is ruled by the following master equation \cite{suppl}:
\begin{align}
\dot{\rho}_{\text{qm}}(t) =\,
& -\frac{\ii}{\hbar}[H_{\text{qm}}, \rho_{\text{qm}}(t)]  \nonumber\\
&+ \gamma \bar{n}_{\omega(\beta_0(t))} D[\sigma^\dagger\otimes \mathbf{1}_{\text{m}}]\rho_{\text{qm}}(t) \nonumber\\
&+ \gamma \left(\bar{n}_{\omega(\beta_0(t))} + 1\right)D[\sigma \otimes \mathbf{1}_{\text{m}}]\rho_{\text{qm}}(t). \label{master_eq}
\end{align}
We have defined the super-operator $D[X]\rho = X\rho X^\dagger - \frac{1}{2}\{X^\dagger X, \rho\}$ and $\sigma= \dyad{g}{e}$.

Product states of the form $\rho_\text{qm}(t) = \rho_\text{q}(t)\otimes  \rho_\text{m}(t)$ are natural solutions of Eq.~\eqref{master_eq}, giving rise to two reduced coupled equations respectively governing the dynamics of the qubit and the mechanics: 
\begin{align}
\dot{\rho}_\text{q}(t) =\,
& -\frac{\ii}{\hbar}[H_{\text{q}}(t), \rho_{\text{q}}(t)]  \nonumber+ \gamma \bar{n}_{\omega(\beta_0(t))} D[\sigma^\dagger]\rho_{\text{q}}(t) \label{ev_q}\\
&+ \gamma \left(\bar{n}_{\omega(\beta_0(t))} + 1\right)D[\sigma]\rho_{\text{q}}(t), \\
\dot{\rho}_\text{m}(t) = & -\frac{\ii}{\hbar}[ H_\text{m}(t) , \rho_{\text{m}}(t)]. \label{ev_mec}
\end{align}
We have introduced the effective time-dependent Hamiltonians: $H_\text{q}(t) = \Tr_\text{m}[\rho_\text{m}(t)(H_\text{q}+V_\text{qm})] = \hbar \omega(\beta_0(t)) \dyad{e}$ and $H_\text{m}(t) = \Tr_\text{q}[\rho_\text{q}(t)(H_\text{m}+V_\text{qm})]$. The physical meaning of these semi-classical equations is transparent: The force exerted by the qubit results into the effective Hamiltonian $H_\text{m}(t)$ ruling the mechanical evolution. Reciprocally, the mechanics modulates the frequency $\omega(\beta_0(t))$ of the qubit (Eq.~\eqref{omega_eff}), which causes the coupling parameters of the qubit to the bath to be time-dependent.

The semi-classical regime of hybrid opto-mechanical systems is especially appealing for quantum thermodynamical purposes, since it allows modeling the time-dependent Hamiltonian ruling the dynamics of a system (the qubit) by coupling this system to a quantum entity, i.e. a quantum battery (the MO). The Hamiltonian of the compound is time-independent, justifying to call it an ``autonomous machine"\cite{frenzel_quasi-autonomous_2016, tonner_autonomous_2005,holmes_coherent_2018}. As demonstrated in a previous work \cite{rev_work_extraction}, this scenery suggests a new strategy to measure average work exchanges in quantum open systems. Defining the average work rate received by the qubit as $\langle \dot{W} \rangle= \Tr_\text{q}[\rho_\text{q}(t)\dot{H}_\text{q}(t)]$, we have shown that this work rate exactly compensates the mechanical energy variation rate: $\langle \dot{{\cal E}}_\text{m} \rangle= \Tr_\text{m}[\dot{\rho}_\text{m}(t){H}_\text{m}] = -\langle \dot{W} \rangle$. Remarkably, this relation demonstrates the possibility of measuring work ``in situ", directly inside the battery. This strategy offers undeniable practical advantages, since it solely requires to measure the mechanical energy at the beginning and at the end of the transformation. The corresponding mechanical energy change is potentially measurable in the ultra-strong coupling regime $g_\text{m}/\Omega \gg 1$ \cite{rev_work_extraction}, which is fully compatible with the semi-classical regime $g_\text{m} t \ll |\beta_0|$.

Our goal is now to extend this strategy to work {\it fluctuations}. A key point is to demonstrate that the qubit and the mechanical state remain in a pure product state along single realizations of the protocol, allowing to unambiguously define stochastic energies for each entity. This calls for an advanced theoretical treatment based on the quantum trajectories picture.\\

\subsection*{Quantum trajectories.}
We shall now describe the evolution of the machine between the time $t_0$ and $t_N$ by stochastic quantum trajectories of pure states $\rightvect{\Sigma} := \{ \ket{\Psi_{\Sigma}(t_n)} \}_{n=0}^N$, where $\ket{\Psi_\Sigma(t_n)}$ is a vector in the Hilbert space of the machine and $t_n = t_0 + n\Delta t$ with $\Delta t$ the time increment. To introduce our approach we first consider the semi-classical regime where the master equation \eqref{master_eq} is valid: The initial state of the machine $\ket{\Psi_{\Sigma}(t_0)}$ is drawn from the product state $\rho_\text{q}(0)\otimes \dyad{\beta_0}$ where $\rho_\text{q}(0)$ is diagonal in the $\{ e,g \}$ basis, and the evolution is studied over a typical duration $(t_N - t_0) \ll |\beta_0| g_\text{m}^{-1}$. Eq.~\eqref{master_eq} is unraveled in the quantum jump picture \cite{plenio_quantum-jump_1998, Gardiner, CarmichaelII, Wisemanbook, haroche}, giving rise to the following set of Kraus operators $\{J_{-1}(t_n);J_{+1}(t_n);J_0(t_n)\}$:
\begin{align} \label{eq:jump}
J_{-1}(t_n) &= \sqrt{\gamma \Dt(\bar{n}_{\omega(\beta_0(t_n))} + 1)}\; \sigma \otimes \mathbf{1}_{\text{m}},\\
J_{+1}(t_n) &= \sqrt{\gamma \Dt\bar{n}_{\omega(\beta_0(t_n))}}\;\sigma^\dagger \otimes \mathbf{1}_{\text{m}},\\
J_0(t_n) &= \mathbf{1}_{\text{qm}} - \frac{\ii\Dt}{\hbar} H_{\text{eff}}(t_n).
\end{align}
We have introduced $\mathbf{1}_{\text{qm}}= \mathbf{1}_{\text{m}} \otimes \mathbf{1}_{\text{q}}$ the identity operator in the Hilbert space of the machine. $J_{-1}$ and $J_{+1}$ are the so-called jump operators. Experimentally, they are signaled by the emission or absorption of a photon in the bath, that corresponds to the transition of the qubit in the ground or excited state respectively. The mechanical state remains unchanged. Reciprocally, the absence of detection event in the bath corresponds the no-jump operator $J_0$, i.e. a continuous, non Hermitian evolution governed by the effective Hamiltonian $H_\text{eff}(t_n) = H_\text{qm} + H_\text{nh}(t_n)$. Here $H_\text{nh}(t) = -(\ii \hbar/ 2) (J_{+1}^\dagger(t) J_{+1}(t) + J_{-1}^\dagger(t) J_{-1}(t))$ is the non-hermitian part of $H_\text{eff}$. 

Let us suppose that the machine is initially prepared in a pure state $\ket{ \Psi(t_0)} = \ket{\epsilon_0,\beta_0}$. The quantum trajectory $\rightvect{\Sigma}$ is then perfectly defined by the sequence of stochastic jumps/no-jump $ \{ {\cal K}_\Sigma(t_n) \}_{n=1}^{N}$ where ${\cal K} = 0,-1,+1$. Namely, $\ket{\Psi_\Sigma(t_N) } = \left(\prod_{n=1}^N J_{{\cal K}_\Sigma(t_n)}  \ket{\Psi(t_0)}\right)/ \sqrt{P[\rightvect{\Sigma}  | \Psi(t_0)] }$ where we have introduced $P[\rightvect{\Sigma} | \Psi(t_0)]  = \prod_{n=1}^{N} \!P[\Psi_\Sigma(t_{n}) | \Psi_\Sigma(t_{n-1}) ]$ the probability of the trajectory $\rightvect{\Sigma}$ conditioned to the initial state $\ket{ \Psi(t_0)}$. $P[\Psi_\Sigma(t_{n}) | \Psi_\Sigma(t_{n-1}) ] = \ev{J^\dagger_{{\cal K}_\Sigma(t_n)}J_{{\cal K}_\Sigma(t_n)}}{\Psi_\Sigma(t_{n-1})}$ denotes the probability of the transition from $\ket{ \Psi_\Sigma(t_{n-1})}$ to $\ket{ \Psi_\Sigma(t_{n})}$ at time $t_{n}$. At any time $t_N$, the density matrix of the machine, i.e. the solution of Eq.~\eqref{master_eq}, can be recovered by averaging over the trajectories: 
\begin{equation} \label{rho_qm}
\rho_\text{qm}(t_N) = \sum_{\rightvect{\Sigma} } P[\rightvect{\Sigma}] \dyad{\Psi_\Sigma(t_N)}.
\end{equation}
We have introduced the probability of the trajectory $P[\rightvect{\Sigma}] = p[\Psi(t_0)] P[\rightvect{\Sigma} | \Psi(t_0)]$, where $p[\Psi(t_0)]$ the probability that the machine is initially prepared in $\ket{\Psi(t_0)}$. 

Interestingly from the expression of the Kraus operators, it appears that starting from the product state $ \ket{\epsilon_0,\beta_0}$, the machine remains in a product state $\ket{ \Psi_\Sigma(t_n)}  = \ket{\epsilon_\Sigma(t_n),\beta_\Sigma(t_n)}$ at any time $t_n$, which is the first result of this paper.  The demonstration is as follows: At each time step $t_n$, either the machine undergoes a quantum jump $J_{\pm 1}$, or it evolves under the no-jump operator $J_0$.  In the former case, the qubit jumps from $\ket{\epsilon_\Sigma(t_n)}$ into $\ket{\epsilon_\Sigma(t_{n+1})}$ and the mechanical state remains unchanged, such as $\ket{\beta_\Sigma(t_{n+1})} = \ket{\beta_\Sigma(t_n)}$. In the latter case, the evolution of the machine state is governed by the effective Hamiltonian $H_\text{eff}$, whose non-hermitian part can be rewritten $H_\text{nh} = (-i\hbar /2) \mathbf{1}_{\text{m}} \otimes H^{q}_\text{nh}$ with $H^{q}_\text{nh}$ diagonal in the bare qubit energy eigenbasis. It naturally derives from the evolution rules that $H_\text{nh}$ has no effect on a machine state of the form $\ket{\epsilon_\Sigma(t_n),\beta_\Sigma(t_n)}$, such that the no-jump evolution reduces to its unitary component defined by $H_\text{qm}$. As studied above, the qubit energy state is stable under such evolution, such that $\ket{\epsilon_\Sigma(t_n)} = \ket{\epsilon_\Sigma(t_{n+1})}$. Reciprocally, the coherent nature of the mechanical field is preserved by $H_\text{m}^{\epsilon_\Sigma(t_n)}$. Thus the mechanics evolves into $\ket{\beta_\Sigma(t_{n+1})} = \exp(-\ii \Dt H_\text{m}^{\epsilon_\Sigma(t_n)}) \ket{\beta_\Sigma(t_n)}$, completing the demonstration. 

This result invites to recast the machine trajectory as a set of two reduced trajectories $\rightvect{\Sigma} = \{ \rightvect{\epsilon}, \rightvect{\beta}[\rightvect{\epsilon}\,]  \}$ where $ \rightvect{\epsilon} = \{ \ket{\epsilon_\Sigma(t_n)} \}_{n=0}^N$ is the stochastic qubit trajectory with $\epsilon_\Sigma(t_n) = e,g$. In the semi-classical regime considered here, the jump probabilities solely depend on $\omega(\beta_0(t))$, such that the qubit reduced evolution is Markovian. Conversely, $\rightvect{\beta} = \{ \ket{\beta_\Sigma(t_n)} \}_{n=0}^N$ is the continuous MO trajectory verifying 
$\ket{\beta_\Sigma(t_n)} = \prod_{k=0}^{n-1} \exp(-\ii  \Dt H_\text{m}^{\epsilon_\Sigma(t_k)})  \ket{\beta_0}$. At any time $t_N$, the mechanical state depends on the complete qubit trajectory $\rightvect{\epsilon}$. 

Examples of numerically generated mechanical trajectories $\rightvect{\beta}[\rightvect{\epsilon}\,]$  (See Methods) are plotted in Fig.~\ref{fig1}c. As it appears in the figure, at the final time the mechanical states $\ket{\beta_\Sigma(t_N)}$ are restricted within an area of typical dimension $\gm/\Omega$. Splitting the mechanical amplitude as $\beta_\Sigma(t_N) = \beta_0(t_N) + \delta \beta_\Sigma(t_N)$, the semi-classical regime is characterized by $\vert \delta \beta_\Sigma(t_n)\vert \ll \vert  \beta_0(t_N)\vert$ while in the ultra-strong coupling regime  $|\delta \beta_\Sigma(t_N)| \gg 1$. These two regimes are compatible, which is the key of our proposal as we show in the next Section.    \\

Interestingly, the modeling of the machine stochastic evolution can be extended over timescales $t \geq |\beta_0|g_\text{m}^{-1}$, beyond the semi-classical regime. The key point is that the trajectory picture allows keeping track of the mechanical state at each time step $\ket{\beta_\Sigma(t_n)}$. Therefore at each time $t_n$, a master equation of the form of Eq.~\eqref{master_eq} can thus be derived and unraveled into a set of {\it trajectory-dependent} Kraus operators similar to Eq.~\eqref{eq:jump}, taking now $\omega(\beta_\Sigma(t_n))$ as the qubit effective frequency.  In this general situation, the machine stochastic evolution still consists in trajectories of pure product states $\ket{\Psi_\Sigma(t_n)} = \ket{\epsilon_\Sigma(t_n), \beta_\Sigma(t_n)}$, but the mechanical fluctuations $| \delta \beta_\Sigma(t_n) |$ cannot be neglected anymore with respect to the mean amplitude $|\beta_0(t_n)|$. Consequently, Eq.\eqref{rho_qm} can not be written as an {\it average} product state of the qubit and the MO, resulting in the emergence of classical correlations between the qubit and the MO average states. Moreover, the jump probabilities at time $t_n$ now depend on $\bar{n}_{\omega(\beta_\Sigma(t_n))}$, such that the reduced qubit trajectory $\rightvect{\epsilon}$ is not Markovian anymore. As we show below, this property conditions the validity of our proposal, which is restricted to the Markovian regime. \\

\subsection*{Stochastic thermodynamics.}
From now on we focus on the following protocol: At the initial time $t_0$ the machine is prepared in a product state $\rho_\text{qm}(t_0) = \rho^\infty_\text{q}(\beta_0) \otimes \dyad{\beta_0}$ where $\rho^\infty_\text{q}(\beta_0)$ is the qubit thermal distribution defined by the effective frequency $\omega(\beta_0)$. Note that $\rho_\text{qm}(t_0)$ is {\it not} an equilibrium state of the whole machine. One performs an energy measurement of the qubit, preparing the state $\ket{\Psi(t_0)} = \ket{\epsilon(t_0), \beta_0}$ with probability $p^\infty_{\beta_0}[\epsilon] = \exp(-\hbar\omega(\beta_0)\delta_{\epsilon,e}/\kT)/Z(\beta_0)$. $Z(\beta_0) = 1 + \exp(-\hbar\omega(\beta_0)/\kT)$ is the partition function. The machine is then coupled to the bath and its evolution is studied between $t_0=0$ and $t_N = \pi/2\Omega$. Depending on the choice of thermodynamical system, this physical situation can be studied from two different perspectives, defining two different transformations. If the considered thermodynamical system is the machine, then the studied evolution corresponds to a relaxation towards thermal equilibrium. Since the machine Hamiltonian $H_\text{qm}$ is time-independent, energy exchanges reduce to heat exchanges between the machine and the bath. On the other hand, if the considered thermodynamical system is the qubit, then the studied transformation consists in driving the qubit out of equilibrium through the time-dependent Hamiltonian $H_\text{q}(t)$, the driving work being provided by the mechanics. In the semi-classical regime, the qubit evolution is Markovian, such that this last situation simply corresponds to Jarzynski's protocol with $H_\text{q}(t) = \hbar \omega(\beta_0(t)) \dyad{e}$. 

We now define and study the stochastic thermodynamical quantities characterizing the transformation experienced by the system (qubit or machine) for the protocol introduced above. 
As shown previously, starting from a product state $\ket{\Psi(t_0)}= \ket{\epsilon_0, \beta_0}$ the machine remains in a product state at any time $\ket{\Psi_\Sigma(t_n)}= \ket{\epsilon_\Sigma(t_n), \beta_\Sigma(t_n)}$.  Defining as ${\cal E}_\text{qm}(\Psi_{\Sigma(t_n)}) = \ev{H_\text{qm}}{\Psi_{\Sigma(t_n)}}$ the machine internal energy, it thus naturally splits into a sum of the qubit energy ${\cal E}_\text{q}(\epsilon_\Sigma(t_n),\beta_\Sigma(t_n))$ (See Eq.~\eqref{Eq}) and mechanical energy ${\cal E}_\text{m}(\beta_\Sigma(t_n))$ (See Eq.~\eqref{Em}).  Along the trajectory, the set of internal energies can change in two distinct ways. A quantum jump taking place at time $t_n$ stochastically changes the qubit and the machine energies by the same amount $\delta {\cal E}_\text{q}[\Sigma,t_n] = \delta {\cal E}_\text{qm}[\Sigma,t_n]$, leaving the MO energy unchanged. Following standard definitions in stochastic thermodynamics \cite{Alicki79,Horowitz12,elouard_role_2017}, the corresponding energy change is identified with heat $q[\Sigma,t_{n}]$ provided by the bath. Conversely in the absence of jump, the qubit remains in the same state between $t_n$ and $t_{n+1}$ while its energy eigenvalues evolve in time due to the qubit-mechanical coupling. Such energy change is identified with work denoted $w[\Sigma,t_n]$ and verifies $ \delta {\cal E}_\text{q}[\Sigma,t_n]  = w[\Sigma,t_n]$. During this time interval, the machine is energetically isolated such that $ \delta {\cal E}_\text{qm}[\Sigma,t_n] = 0$. Therefore the work increment exactly compensates the mechanical energy change $\delta {\cal E}_\text{m}[\Sigma,t_n] = -w[\Sigma,t_n]$. Finally, the total work (resp. heat) received by the qubit is defined as  $W\Traj = \sum_{n = 0}^{N - 1} w[\Sigma,t_n]$ (resp. $Q\Traj = \sum_{n = 0}^{N - 1} q[\Sigma,t_n] $). By construction, their sum equals the qubit total energy change between $t_0$ and $t_N$, $\Delta {\cal E}_\text{q}\Traj = W\Traj + Q\Traj$. From the analysis conducted above, it appears that the heat exchange corresponds to the energy change of the machine, $\Delta {\cal E}_\text{qm}\Traj = Q\Traj$. Reciprocally, the work received by the qubit is entirely provided by the mechanics and verifies:

\begin{equation}
W\Traj = -\Delta {\cal E}_\text{m}\Traj, \label{autonomous}
\end{equation}
which is the second result of this article. Eq.~\eqref{autonomous} extends the results obtained for the average work in a previous work \cite{rev_work_extraction}, and explicitly demonstrates the one-by-one correspondence between the stochastic work received by the qubit and the mechanical energy change between the start and the end of the trajectory. The MO thus behaves as an ideal embedded quantum work meter at the single trajectory level. \\

We finally derive the expression of the stochastic entropy production $\ds\Traj$. It is defined by comparing the probability of the forward trajectory in the direct protocol $P\Traj$ to the probability of the
backward trajectory in the time-reversed protocol $\tilde{P}\rTraj$\cite{thermo_traj_Broeck}: 
\begin{equation} \label{entropy}
\ds\Traj = \log \left( \frac{P\Traj}{\tilde{P}\rTraj} \right).
\end{equation}

The probability of the direct trajectory reads:
\begin{equation}
P\Traj = p^\infty_{\beta_0}[\epsilon_\Sigma(t_0)]\prod_{n=1}^{N} \!P[\Psi_\Sigma(t_{n}) | \Psi_\Sigma(t_{n-1}) ], \label{Pd}
\end{equation}
The state of the hybrid system averaged over the forward trajectories at time $t_N$ is described by Eq.~\eqref{rho_qm}. At the end of the protocol, the reduced mechanical average state defined as $\rho_\text{m} (t_N) = \text{Tr}_\text{q} [\rho_\text{qm} (t_N)]$ thus consists in a discrete distribution of the final mechanical states $\{ \ket{\beta_\Sigma(t_N)} \}$. Introducing the probability $p_\text{m}[\beta_\text{f}]$ for the mechanical amplitude to end up in a state of amplitude $\beta_\text{f}$, we shall denote it in the following $\rho_\text{m} (t_N) =  \Sigma_{\beta_\text{f}} p_\text{m}[\beta_\text{f}] \dyad{\beta_\text{f}}$ where $\Sigma_{\beta_\text{f}} p_\text{m}[\beta_\text{f}] = 1$. \\

Reciprocally, the time-reversed protocol is defined between $t_N$ and $t_0$. It consists in time-reversing the unitary evolution governing the dynamics of the machine, keeping the same stochastic map at each time $t_n$. This leads to the expression of the time dependent reversed Kraus operators\cite{crooks_quantum_2008, elouard_role_2017, Manzano17, Manikandan18}: 
\begin{align}
\tilde{J}_0(t_n)& = \mathbf{1}_{\text{qm}} + \frac{\ii\Dt}{\hbar} H^\dagger_\text{eff}(t_n),\\
\tilde{J}_{-1}(t_n) & = J_{+1}(t_n),\\
\tilde{J}_{+1}(t_n) &= J_{-1}(t_n), 
\end{align}

The initial state of the backward trajectory is defined as follows: The mechanical state $\ket{\beta_\Sigma(t_N)}$ is drawn from the final distribution of states $\{ \ket{\beta_\text{f}} \}$ generated by the direct protocol with probability $p_\text{m}[\beta_\text{f}]$, while the qubit state is drawn from the thermal equilibrium defined by $\beta_\Sigma(t_N)$ with probability $p^\infty_{\beta_\Sigma(t_N)}$. The probability of the backward trajectory reads
\begin{align}
\tilde{P}\rTraj =\,&  p_\text{m}[\beta_\Sigma(t_N)] p^\infty_{\beta_\Sigma(t_N)}[\epsilon_\Sigma(t_N)] \nonumber\\
&\quad\times\prod_{n = N}^{1}\tilde{P}[\Psi_\Sigma(t_{n-1}) | \Psi_\Sigma(t_{n}) ]. \label{Pr}
\end{align}
We have introduced the reversed jump probability at time $t_n$ $\tilde{P}[\Psi_\Sigma(t_{n-1}) | \Psi_\Sigma(t_{n}) ] =\ev {\tilde{J}^\dagger_{{\cal K}_\Sigma(t_n)} \tilde{J}_{{\cal K}_\Sigma(t_n)} }{\Psi_\Sigma(t_{n}) }$. Based on Eqs.~ \eqref{autonomous}, \eqref{entropy}, \eqref{Pd}, \eqref{Pr}, we derive in the Supplementary\cite{suppl} the following expression for the stochastic entropy produced along $\rightvect{\Sigma}$:
\begin{equation} \label{Si_machine}
\ds\Traj = \sigma\Traj + I_{\text{Sh}}\Traj,
\end{equation}
where $\sigma\Traj$ and  $I_{\text{Sh}}\Traj$ are defined as
\begin{align}
\sigma \Traj  & =  -\frac{\Delta {\cal E}_\text{m} \Traj + \Delta F \Traj}{k_\text{B}T}, \label{sigma_q}\\
I_{\text{Sh}}\Traj & = -\log( p_\text{m}[\beta_\Sigma(t_N)]). \label{sigma_m}
\end{align}

We have introduced the quantity $\Delta F \Traj = k_\text{B} T \log(Z(\beta_0)/Z (\beta_\Sigma(t_N)))$ that extends the notion of the qubit free energy change to cases where the reduced qubit trajectory $\rightvect{\epsilon}$ is non-Markovian. In the Markovian regime, we simply recover $Z(t_N) =1 + \exp(-\hbar \omega(\beta_0(t_N))/ k_\text{B}T)$ and $\Delta F\Traj = \Delta F$.  As we show below, in this case $\sigma\Traj$ can be interpreted as the entropy produced along the reduced trajectory of the qubit, that gives rise to a reduced JE. Conversely, $I_{\text{Sh}}\Traj $ measures the stochastic entropy increase of the MO and is involved in a generalized IFT characterizing the evolution of the whole machine. We now study in detail these two fluctuation theorems.\\

\begin{figure}[htb!]
    \includegraphics[scale=1]{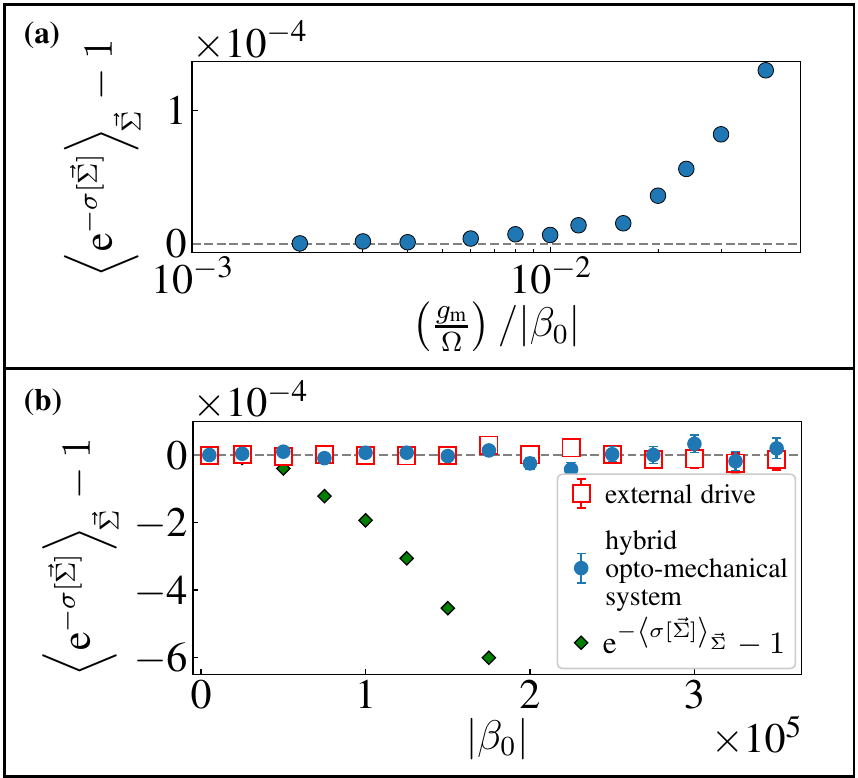}
    \caption{Jarzynski's equality for the qubit. Parameters: $T = 80$ K, $\hbar\omega_0 = 1.2\kT$, $\Omega/2\pi = 100$ kHz, $\gamma/\Omega = 5$. \textbf{(a)} Deviation from JE as a function of $\left(\frac{\gm}{\Omega}\right)/|\beta_0|$ ($|\beta_0| = 5000$). The points were computed by increasing the opto-mechanical coupling strength $\gm/2\pi$ from 1 MHz to 20 MHz, keeping the other parameters constant.  
        \textbf{(b)} Deviation from JE as a function of $|\beta_0|$ with $\gm/\Omega = 10$. Red squares: Case of a classical external drive imposing the qubit frequency modulation $\omega(\beta_0(t))$ (See text). Blue dots:  Eq.~\eqref{JE}. Green diamonds: $\exp(-\ev{\sigma\Traj}_{\rightvect{\Sigma}}) - 1$. These green points demonstrate that JE is not trivially reached because the considered transformations are reversible. 
        \label{fig2}}
\end{figure}

\subsection*{Reduced Jarzynski's equality.}
We first focus on the transformation experienced by the qubit. As mentioned above, in the Markovian regime the applied protocol corresponds to Jarzynski's protocol: The qubit is driven out of thermal equilibrium while it experiences the frequency modulation $\omega(\beta_0(t))$. Since the stochastic work $W\Traj$ is provided by the mechanics, one expects the mechanical energy fluctuations to obey a reduced Jarzynski's equality. We derive in the Supplementary \cite{suppl} the following IFT:

\begin{equation}
\ev{\exp(\frac{\Delta {\cal E}_\text{m} \Traj} {k_\text{B}T})} _{\rightvect{\Sigma}}= \exp(-\frac{\Delta F} {k_\text{B}T}).  \label{JE}
\end{equation}

Eq.~\eqref{JE} corresponds to the usual Jarzynski's equality, with the remarkable difference that the stochastic work involved in $\sigma\Traj$ is now replaced by the mechanical energy change $\Delta {\cal E}_\text{m}\Traj$. This is the third and most important result of this paper, which now suggests a new strategy to measure work {\it fluctuations}. Instead of reconstructing the stochastic work by monitoring the complete qubit trajectory, one can simply measure the mechanical stochastic energy at the beginning and at the end of the protocol. This can be done, e.g. in time-resolved measurements of the mechanical complex amplitude through to optical deflection techniques\cite{Sanii10,Mercier16}. To do so, the final mechanical states $\ket{\beta_\Sigma(t_N)}$ should be distinguishable, which requires to reach the ultra-strong coupling regime. As mentioned above, this regime has been experimentally evidenced \cite{trompette} with typical values $\Omega \sim \gm \sim 400$kHz. The strategy we suggest here is drastically different from former proposals aiming at measuring JE in a quantum open system, that involved challenging reservoir engineering techniques \cite{Horowitz12,elouard_probing_2017} or fine thermometry \cite{pekola} in order to measure heat exchanges. 

\begin{figure}[htb]
    \includegraphics[scale=1]{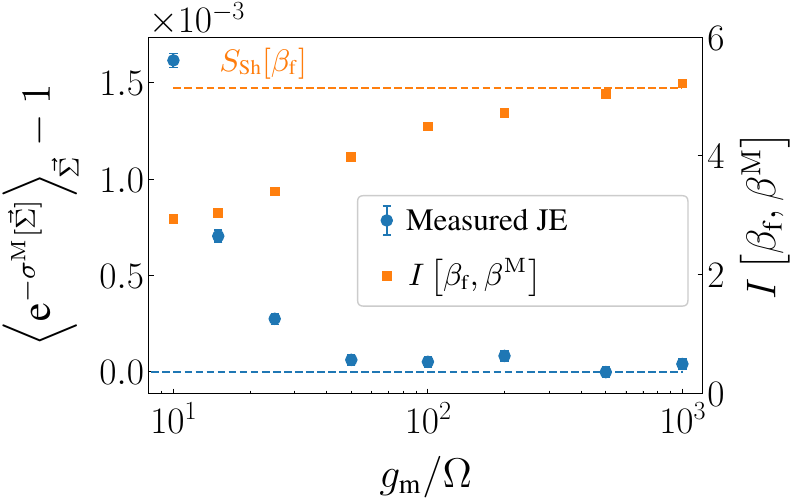} 
    \caption{Impact of finite precision readout of the mechanical amplitude. Parameters: $\delta\beta = 2$, $T = 80$ K, $\hbar\omega_0 = 1.2\kT$, $\Omega/2\pi = 1$ kHz, $\gamma/\Omega = 5$. $2 \gm|\beta_0|$ was kept constant ($2 \gm|\beta_0|/2\pi = 600$ GHz) while increasing $\gm$, such that each point corresponds to the same mean reduced entropy production $\langle \sigma{\Traj} \rangle_{\rightvect{\Sigma}}$. Left axis, blue dots: Deviation from measured JE. Right axis, orange squares: Mutual information $I[\beta_\text{f}, \beta^\text{M}]$. Orange dashed line: Shannon's entropy of the final distribution of mechanical states $S_\text{Sh}[\beta_\text{f}]$ (See text).\label{fig3}}
\end{figure}

We have simulated the reduced JE (See Fig.~\ref{fig2}a).  As expected, JE is verified in the Markovian limit where we have checked that the action of the MO is similar to a classical external operator imposing the qubit frequency modulation $\omega(\beta_0(t))$ (Fig.~\ref{fig2}b). On the contrary, the Markovian approximation and JE break down in the regime $(\gm/\Omega)/|\beta_0|  \geq 10^{-2}$. In what follows, we restrict the study to the range of parameters $(\gm/\Omega)/|\beta_0| < 10^{-2}$.\\

The results presented in Fig.~\ref{fig2} presuppose the experimental ability to measure the mechanical states with an infinite precision. To take into account both quantum uncertainty and experimental limitations, we now assume that the measured complex amplitude $\beta^\text{M}$ corresponds to the mechanical amplitude $\beta_\text{f}$ in the end of the protocol with a finite precision $\delta \beta$. For our simulations we have chosen $\delta \beta = 2$ which corresponds to achievable experimental value \cite{Sanii10,Mercier16}. To quantify this finite precision, we introduce the mutual information between the final distribution of mechanical states $p_\text{m}[\beta_\text{f}]$ introduced above, and the measured distribution $p_\text{m}[\beta^\text{M}]$, defined as:
\begin{align}
   & I[\beta_\text{f}, \beta^\text{M}] = \nonumber\\
   & \quad\sum_{\beta_\text{f}, \beta^\text{M}}p(\beta_\text{f}, \beta^\text{M}) \log(\frac{p(\beta_\text{f}, \beta^\text{M})}{p_\text{m}[\beta_\text{f}] p_\text{m}[\beta^\text{M}]}).
\end{align}
$p(\beta_\text{f}, \beta^\text{M})$ denotes the joint probability of measuring $\beta^\text{M}$ while the mechanical amplitude equals $\beta_\text{f}$.
If the measurement precision is infinite, the mutual information $I[\beta_\text{f}, \beta^\text{M}]$ exactly matches the Shannon entropy characterizing the final distribution of mechanical states $S_\text{Sh}[\beta_\text{f}] = -\sum_{\beta_\text{f}} p_\text{m}[\beta_\text{f}] \log(p_\text{m}[\beta_\text{f}])$.  On the opposite, it vanishes in the absence of correlations between the two distributions.

The simulation of the measured JE and the mutual information $I[\beta_\text{f}, \beta^\text{M}]$ are plotted in Fig.~\ref{fig3} for the measurement precision $\delta\beta = 2$, as a function of the parameter $g_\text{m}/\Omega$ (See Methods). We have introduced the measured reduced entropy production $\sigma^\text{M}\Traj = (W^M\Traj - \Delta F)/ k_\mathrm{B} T$ where $W^M\Traj$ is the measured work distribution $W^M\Traj = -\Delta {\cal E}_\text{m}^\text{M}\Traj = \hbar\Omega(|\beta_0^\text{M}|^2 - |\beta^\text{M}_\Sigma(t_N)|^2)$. As expected, small values of $\gm/\Omega$ correspond to a poor ability to distinguish between the different final mechanical states, hence to measure work, which is characterized by a non-optimal mutual information. In this limit, the measured work fluctuations $W^M\Traj$ do not verify JE. Increasing the ratio $\gm/\Omega$ allows to increase the information extracted on the work distribution during the readout. Thus the mutual information converges towards $S_\text{Sh}[\beta_\text{f}]$ despite the finite precision readout. JE is recovered for $\gm/\Omega \sim 50$.  Such high rates are within experimental reach, by engineering modes of lower mechanical frequency\cite{trompette_suppl}.  \\

\subsection*{Generalized integral fluctuation theorem.}

\begin{figure}[h]
    \includegraphics[scale=1]{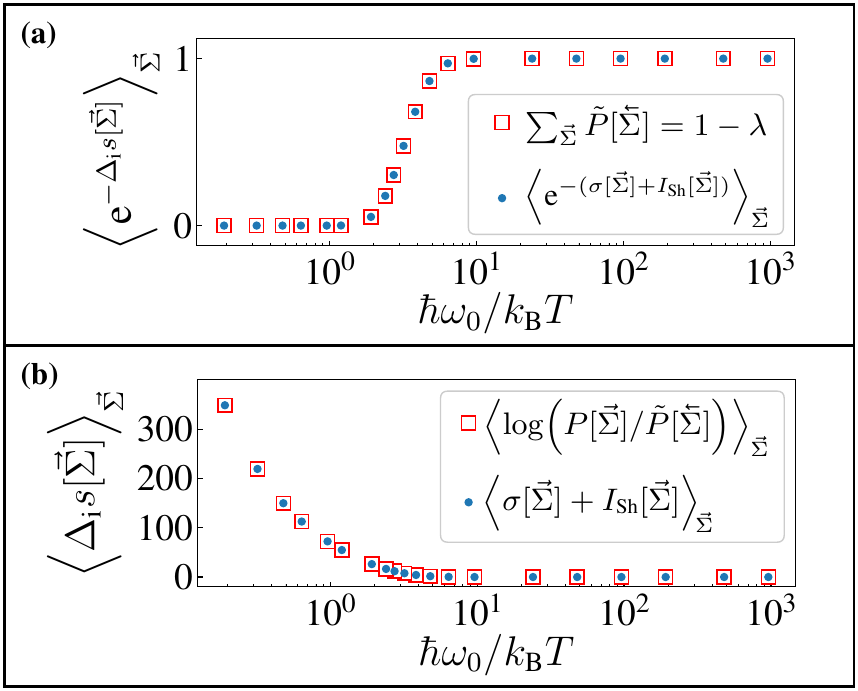}        
    \caption{ \label{fig4}
       Deviation from the integral fluctuation theorem \textbf{(a)}  and mean entropy production \textbf{(b)}  for the complete autonomous machine. Parameters: $\omega_0/ 2\pi = 2$ THz (amounts to $\hbar \omega_0 / \kT = 1.2$ for $T = 80$ K used in Fig.~\ref{fig2}), $\Omega/2\pi = 100$ kHz, $\gamma/\Omega = 5$, $\gm/\Omega= 10$ and $|\beta_0| = 5000$. In both cases, two different expressions were used. The blue dots are computed using the final distribution of mechanical states $\{ \ket{\beta_\Sigma(t_N)} \}$ and mimic an experiment. The red squares involve the probability of the reversed trajectory, which can only be the result of a theoretical treatment. (See Methods for more details.)}
\end{figure}

We finally consider the complete machine as the thermodynamical system under study. Based on Eq.~\eqref{entropy} and \eqref{Si_machine}, we show in the Supplementary \cite{suppl} that the entropy produced along the stochastic evolution of the hybrid system obeys a modified IFT of the form:
\begin{equation}
\ev{\exp(-\Delta_\text{i} s\Traj)}_{\rightvect{\Sigma}} = 1- \lambda. \label{GFT}
\end{equation}
Following \cite{murashita_nonequilibrium_2014,funo2015,Ueda_PRA,Nakamura_arXiv}, we have defined the parameter $\lambda$ as $\sum_{\rightvect{\Sigma}} \tilde{P} \rTraj  = 1 -\lambda$.
The case $\lambda \neq 0$ signals the existence of backward trajectories $\leftvect{\Sigma}$ without any forward counterpart, i.e. $P[\rightvect{\Sigma}] = 0$, a phenomenon that has been dubbed absolute irreversibility (See Supplementary \cite{suppl}). From Eq.~\eqref{GFT} and the convexity of the exponential, it is clear that absolute irreversibility characterizes transformations associated to a strictly positive entropy production. This is the case in the present situation, which describes the relaxation of the machine towards a thermal equilibrium state: Such transformation is never reversible, unless for $T = 0$.

The IFT  (Eq.~\eqref{GFT}) and the mean entropy production $\ev{\ds\Traj}_{\rightvect{\Sigma}}$ are plotted in Fig.~\ref{fig4}a and Fig.~\ref{fig4}b respectively, as a function of the bath temperature $T$ (See Methods). The limit $\hbar\omega_0 \gg \kT$ corresponds to the trivial case of a single reversible trajectory characterized by a null entropy production and $\lambda \rightarrow 0$. In the opposite regime defined by $\kT \gg \hbar \omega_0$, a mean entropy is produced while $\lambda \rightarrow 1$: In this situation, most backward trajectories have no forward counterpart. As we show in the Supplementary\cite{suppl}, such effect arises since a given $\beta_\text{f}$ of the final distribution of mechanical states can only be reached by a single forward trajectory, while it provides a starting point for a large number of backward trajectories. 

As noticed in\cite{murashita_nonequilibrium_2014,Nakamura_arXiv,Manikandan18}, absolute irreversibility can also appear in IFTs characterizing the entropy produced by a measurement process. In particular, $\lambda \neq 0$ can signal a perfect information extraction: This typically corresponds to the present situation which describes the creation of classical correlations between the qubit reduced trajectory $\rightvect{\epsilon}$ and the distributions of final mechanical states $\rightvect{\beta}[\rightvect{\epsilon}]$. Interestingly, the two FTs \eqref{JE} and \eqref{GFT} are thus deeply related. To be experimentally checked, Eq.~\eqref{JE} requires the MO to behave as a perfect quantum work meter, which is signaled by absolute irreversibility Eq.~\eqref{GFT}. Therefore absolute irreversibility is constitutive of the protocol, and a witness of its success.

\section*{Discussion}
\noindent We have evidenced a new protocol to measure stochastic entropy production and thermodynamic time arrow in a quantum open system. Based on the direct readout of stochastic work exchanges within an autonomous machine, this protocol is experimentally feasible in state-of-the-art opto-mechanical devices and robust against finite precision measurements. It offers a promising alternative to former proposals relying on the readout of stochastic heat exchanges within engineered reservoirs, which require high efficiency measurements. Originally, our proposal sheds new light on absolute irreversibility, which quantifies information extraction within the quantum work meter and therefore signals the success of the protocol.\\

In the near future, direct work measurement may become extremely useful to investigate genuinely quantum situations where a battery coherently drives a quantum open system into coherent superpositions. Such situations are especially appealing for quantum thermodynamics since they lead to entropy production and energetic fluctuations of quantum nature \cite{elouard_role_2017,elouard_extracting_2017}, related to the erasure of quantum coherences \cite{Santos2017a,Francica2017}. Recently, small amounts of average work have been directly measured, by monitoring the resonant field coherently driving a superconducting qubit \cite{Cottet7561}. 
Generalizing our formalism to this experimental situation would relate measurable work fluctuations to quantum entropy production, opening a new chapter in the study of quantum fluctuation theorems.

\section*{Methods}

\noindent The numerical results presented in this article were obtained using the jump and no-jump probabilities to sample the ensemble of possible direct trajectories \cite{haroche}. The average value of a quantity $A\Traj$ is then approximated by $\ev{A\Traj}_{\srightvect{\Sigma}} \simeq \frac{1}{N_\text{traj}}\sum_{i=1}^{N_\text{traj}} A[\rightvect{\Sigma}_i]$ where $N_\text{traj} = 5\times 10^6$ is the number of numerically generated trajectories and $\rightvect{\Sigma}_i$ denotes the $i$-th trajectory. 

The reduced entropy production $\sigma\Traj$ used in Fig.~\ref{fig2} and \ref{fig4} was calculated with the expression \eqref{sigma_q}, using the numerically generated values of $\beta_0$ and $\beta_\Sigma(t_N)$ in the trajectory $\rightvect{\Sigma}$.
One value of $\beta_\Sigma(t_N)$ can be generated by a single direct trajectory $\rightvect{\Sigma}$: Below we use the equality $p_\text{m}[\beta_\Sigma(t_N)] = P\Traj$. Using the expression \eqref{Pr} of the probability of the reversed trajectory, the average entropy production becomes:
\begin{align*}
\ev{\Delta_{\text{i}}s\Traj}_{\!\srightvect{\Sigma}} 
&=\ev{\log(\frac{P\Traj}{\tilde{P}\rTraj})}_{\srightvect{\Sigma}}\\
&=\left\langle-\log\left(p^\infty_{\beta_\Sigma(t_N)}[\epsilon_\Sigma(t_N)]\phantom{\prod_{n = 1}^{N}}\right.\right. \\
&\hspace{2cm}\left.\left.\times\prod_{n = 1}^{N}  \tilde{P}[\Psi_\Sigma(t_{n-1}) | \Psi_\Sigma(t_{n}) ]\right)\right\rangle_{\srightvect{\Sigma}} \\
 &\simeq \frac{-1}{N_\text{traj}}\sum_{i = 1}^{N_\text{traj}}\log\left(p^\infty_{\beta^i_\Sigma(t_N)}[\epsilon^i(t_N)]\phantom{\prod_{n = 1}^{N}}\right.\\
 &\hspace{2.7cm}\left.\times\prod_{n = 1}^{N } \tilde{P}[\Psi^i_\Sigma(t_{n-1}) | \Psi^i_\Sigma(t_{n}) ]\right),
\end{align*}
and,
\begin{align*}
\sum_{\srightvect{\Sigma}}\tilde P\rTraj 
 &= \sum_{\srightvect{\Sigma}}  p^\infty_{\beta_\Sigma(t_N)}[\epsilon_\Sigma(t_N)] p_\text{m}[\beta_\Sigma(t_N)]\\ &\hspace{0.9cm}\times\prod_{n = 1}^{N}  \tilde{P}[\Psi_\Sigma(t_{n-1}) | \Psi_\Sigma(t_{n}) ]\\
 &= \ev{ p^\infty_{\beta_\Sigma(t_N)}[\epsilon_\Sigma(t_N)] \prod_{n = 1}^{N} \tilde{P}[\Psi_\Sigma(t_{n-1}) | \Psi_\Sigma(t_{n}) ]}_{\srightvect{\Sigma}}\\
 &\simeq \frac{1}{N_\text{traj}}\sum_{i = 1}^{N_\text{traj}}  p^\infty_{\beta^i_\Sigma(t_N)}[\epsilon^i(t_N)]\\
 &\hspace{1.7cm}\times\prod_{n = 1}^{N} \tilde{P}[\Psi^i_\Sigma(t_{n-1}) | \Psi^i_\Sigma(t_{n}) ].
\end{align*}
The plotted error bars represent the statistical error $\sigma/\sqrt{N_\text{traj}}$, where $\sigma$ is the standard deviation.\\

To obtain Fig.~\ref{fig3}, we considered that the preparation of the initial MO state was not perfect. So instead of starting from exactly $\ket{\beta_0}$, the MO trajectories start from $\ket{\beta_\Sigma(t_0)}$ with the $\beta_\Sigma(t_0)$ uniformly distributed in a square of width $2\delta \beta$, centered on $\beta_0$. Similarly, the measuring apparatus has a finite precision, modeled by a grid of cell width $2\delta\beta$ in the phase plane $(\Re\beta_\text{f}, \Im\beta_\text{f})$. Instead of obtaining the exact value of $\beta_\Sigma(t_N)$, we get $\beta^\text{M}_\Sigma(t_N)$, the center of the grid cell in which $\beta_\Sigma(t_N)$ is.
The value used to compute the thermodynamical quantities are not the exact $\beta_\Sigma(t_0)$ and $\beta_\Sigma(t_N)$ but $\beta^\text{M}_0 = \beta_0$ and $\beta^\text{M}_\Sigma(t_N)$.

\section*{Acknowledgment} 
\noindent J.M. acknowledges J-P Aguilar Ph.D. grant from the CFM foundation. C.E. acknowledges the US Department of Energy grant No. de-5sc0017890. This work was supported by
the ANR project QDOT (ANR-16-CE09-0010-01). Part of this work was discussed at the Kavli Institute for Theoretical Physics during the program Thermodynamics of quantum systems: Measurement, engines, and control. The authors acknowledge the National Science Foundation under Grant No. NSF PHY-1748958.


\onecolumngrid
\newpage
\section*{\large Supplementary information for ``An autonomous quantum machine to measure the thermodynamic arrow of time''}
\vspace*{0.5cm}
\supplsection{Master equation.} 
Here we describe the coupling of a hybrid opto-mechanical system to a thermal bath of temperature $T$.
The total Hamiltonian reads $H = H_\text{qm} + H_\text{b} + V_\text{qb}$, where $H_{\text{b}} = \sum_{k}\hbar \omega_k a^\dagger_k a_k$ is the free Hamiltonian of the bath, $a_k$ is the annihilation operator of the $k$-th electromagnetic mode of frequency $\omega_k$. The coupling Hamiltonian between the qubit and the bath 
in the Rotating Wave Approximation equals $V_{\text{qb}} = \sum_k \hbar g_k (a_k \sigma^\dagger + a^\dagger_k \sigma)$
, where $\sigma= \dyad{g}{e}$ and $g_k$ is the coupling strength between the qubit and the $k$-th mode. We denote $\gamma =  \sum_k g_k^2\delta(\omega_0-\omega_k)$ the spontaneous emission rate of the bare qubit in the bath. The typical correlation time of the bath verifies $\tau_{\text{c}} \ll \gamma^{-1},  g_{\text{m}}^{-1}, \Omega^{-1}$.

The hybrid system is initially prepared in a factorized state $\rho_\text{qm}(0) = \rho_\text{q}(0)\otimes \dyad{\beta_0}$ where $\rho_\text{q}(0)$ is diagonal in the bare qubit energy basis and $\ket{\beta_0}$ is a  pure coherent state. We can define a coarse grained time step $\Dt$, fulfilling  $\tau_{\text{c}} \ll \Dt \ll \gamma^{-1}$ such that under these assumptions, the hybrid system and the bath are always in a factorized state (Born-Markov approximation). Moreover, the coupling to the bath solely induces transitions between the qubit bare energy states, such that the hybrid system naturally evolves into a classically correlated state of the form $\rho_\text{qm}(t) = P_e(t) \dyad{e}\otimes \dyad{\beta_e(t)} + P_g(t) \dyad{g}\otimes \dyad{\beta_g(t)}$. $\{\ket{\beta_{\epsilon}(t)}\}_{\epsilon=e,g}$ are coherent states of the MO verifying $\beta_{\epsilon} = \beta_0 e^{-\ii\Omega t} + \delta \beta_{\epsilon}(t)$. The mechanical fluctuations after a typical time $t$ verify $|\delta \beta_{\epsilon}(t)| \sim g_\text{m}t$. They become potentially detectable as soon as $|\delta \beta_{\epsilon}(t)| \geq 1$, i.e. $t \geq  \gm^{-1}$. Conversely, the mechanical fluctuations have no influence on the qubit frequency as long as $|\delta \beta_{\epsilon}(t)| \ll |\beta_0|$, i.e. $t \ll |\beta_0|\gm^{-1}$ (See main text).

The precursor of the master equation reads 
\begin{align*}
\Delta \rqm^\I(t) = \rqm^\I(t+\Dt) - \rqm^\I(t) =  -\frac{1}{\hbar^2}\int_t^{t+\Delta t}\dd t' \int_t^{t'}\dd t'' \Tr_\text{b}\left[ \left[V_{\text{qb}}^\I(t'), \left[V_{\text{qb}}^\I(t''), \rqm^\I(t)\otimes\rho_\text{b}\right]\right]\right],
\end{align*}
where we have defined the interaction representation with respect to the free Hamiltonians of the hybrid system and the bath $\rqm^\I(t) =  \e^{\ii t (H_\text{qm} + H_\text{b})/\hbar}\rqm(t)\e^{-\ii t (H_\text{qm} + H_\text{b})/\hbar}$, $V_{\text{qb}}^\I(t) =  \e^{\ii t (H_\text{qm} + H_\text{b})/\hbar}V_{\text{qb}}(t)\e^{-\ii t  (H_\text{qm} + H_\text{b})/\hbar}$.  $\Tr_\text{b}$ is the trace over the bath's Hilbert space and $\rho_\text{b}$ is the bath's density matrix.  We have used that the term of first order in $V^\I_\text{qb}$ vanishes. 

Because of the presence of $V_\text{qm} = \hbar \gm \dyad{e}(b + b^\dagger)$ in $H_0$, $V_{\text{qb}}^\I$ also acts on the MO. It can be split in the following way: $V_{\text{qb}}^\I(u)  = R_{\text{b}}^\dagger(u)\otimes S(u) + R_{\text{b}}(u)\otimes S^\dagger(u)$, with $R_{\text{b}}(u) = \hbar \sum_k g_k a_k \e^{-\ii \omega_k u}$ and $S(u) = \e^{\ii u H_\text{qm}/\hbar}(\sigma \otimes \mathbf{1}_\text{m})\e^{-\ii u H_\text{qm}/\hbar}$. Then, expanding the commutators, the trace over the bath's degrees of freedom can be computed. For any two times $u$ and $v$, the correlation functions of the bath read: $\Tr_\text{b}[\rho_{\text{b}} R_\text{b}(u)R_\text{b}(v)] = \Tr_\text{b}[\rho_{\text{b}} R^\dagger_\text{b}(u)R^\dagger_\text{b}(v)] = 0$, $g_-(u,v) = \Tr_\text{b}[\rho_{\text{b}} R_\text{b}(u)R^\dagger_\text{b}(v)] = \hbar^2 \sum_k g_k^2(\bar{n}_{\omega_k} + 1)\e^{-\ii \omega_k (u - v)}$ and $g_+(u,v) = \Tr_\text{b}[\rho_{\text{b}} R^\dagger_\text{b}(u)R_\text{b}(v)] = \hbar^2 \sum_k g_k^2\bar{n}_{\omega_k} \e^{\ii \omega_k (u - v)}$. $\bar{n}_{\omega_k}$ is the average number of photon of frequency $\omega_k$ in the bath. As a result, only terms containing one $S$ and one $S^\dagger$ remains in $\Delta \rqm^\I$. The integral $\int_t^{t'}\dd t''$ can then be changed into an integral over $\tau = t' - t''$: $\int_0^{t' - t}\dd \tau$. Since $g_{s}(u,v) = g_{s}(u - v)$, with $s \in \{+, -\}$, is non zero only for $|u - v| \lesssim \tau_c \ll \Dt$, the upper bound can be set to infinity. In addition, the coarse-graining time can be chosen such that $\Dt \ll \gamma^{-1}, g_\text{m}^{-1}, \Omega^{-1}$ and the MO does not evolve during the integration. As a consequence, the operator $S(u)$ becomes $S(u) = \sigma\e^{-\ii \omega_0 u}\e^{-\ii \gm(b + b^\dagger) u}$. 

As long as $u \ll |\beta_0|\gm^{-1}$, $\omega(\beta_e(u)) \simeq \omega(\beta_g(u))\simeq \omega(\beta_0(u)) $ where $\omega(\beta) = \omega_0 + g_\text{m}(\beta + \beta^*)$ and $\beta_0(t) = \beta_0 e^{-i\Omega t}$. Moreover, the state of the system $\rqm(t)$ can be approximated by the factorized state $\rho_{\text{q}}(t)\otimes\dyad{\beta_0(t)}$. Denoting $\ket{E(u)} = \dyad{e}\otimes \dyad{\beta_0(u)}$ (resp. $\ket{G(u)} = \dyad{g}\otimes \dyad{\beta_0}(u)$), $S(u)$  thus verifies $\ev{S(u)}{E(u)} = \ev{S(u)}{G(u)} = \mel{E(u)}{S(u)}{G(u)} = 0$ and $\mel{G(u)}{S(u)}{E(u)} \simeq e^{-\ii \omega(\beta_0(u)) u}$. $\Delta\rho_{\text{qm}}^\text{I}(t)$ can then be decomposed over the states $\ket{E(t)}$, $\ket{G(t)}$ and the integral over $\tau$ make the system interacts only with bath photons of frequency $\omega(\beta_0(t))$.

As announced in the main text, the master equation describing the relaxation of the hybrid system in the bath can finally be written as 
\begin{equation}
\dot{\rho}_{\text{qm}}(t) =\,
-\frac{\ii}{\hbar}[H_{\text{qm}}, \rho_{\text{qm}}(t)]  + \gamma \bar{n}_{\omega(\beta_{0}(t))} D[\sigma^\dagger\otimes \mathbf{1}_{\text{m}}]\rho_{\text{qm}}(t) + \gamma \left(\bar{n}_{\omega(\beta_{0}(t))} + 1\right)D[\sigma\otimes \mathbf{1}_{\text{m}}]\rho_{\text{qm}}(t), \label{suppl:master_eq}
\end{equation}
where $D[X]\rho = X\rho X^\dagger - \frac{1}{2}\{X^\dagger X, \rho\}$, and $\bar{n}_{\omega} = \left(\exp(\hbar\omega/\kT )- 1\right)^{-1}$.

\supplsection{Entropy production for the autonomous machine.}
Starting from the definition $\ds\Traj = \log(P\Traj/\tilde{P}\rTraj)$ and using Eqs.~(13) and (17) from the main text, the entropy production can be written:
\begin{equation}
\ds\Traj = \log\left(\frac{p^\infty_{\beta_0}[\epsilon_\Sigma(t_0)]}{p_\text{m}[\beta_\Sigma(t_N)]p^\infty_{\beta_\Sigma(t_N)}[\epsilon_\Sigma(t_N)]}
\frac{\prod_{n=1}^{N} P[\Psi_\Sigma(t_{n}) | \Psi_\Sigma(t_{n-1}) ]}{ \prod_{n = 1}^{N}\tilde{P}[\Psi_\Sigma(t_{n-1}) | \Psi_\Sigma(t_{n}) ]}\right).
\end{equation}
From the expressions of the jump and no-jump operators, we obtain
\begin{equation}
\frac{P[\Psi_\Sigma(t_{n}) | \Psi_\Sigma(t_{n-1}) ]}{\tilde{P}[\Psi_\Sigma(t_{n-1}) | \Psi_\Sigma(t_{n}) ]} 
=\frac{\ev{J^\dagger_{{\cal K}_\Sigma(t_n)}J_{{\cal K}_\Sigma(t_n)}}{\Psi_\Sigma(t_{n-1})}}{\ev{\tilde{J}^\dagger_{{\cal K}_\Sigma(t_n)}\tilde{J}_{{\cal K}_\Sigma(t_n)}}{\Psi_\Sigma(t_{n})}} =\exp(-q[\Sigma, t_{n-1}]/\kT), \\
\end{equation}
and, using the expression of the thermal distribution  $p^\infty_{\beta}[\epsilon] = \exp(-\hbar\omega(\beta)\delta_{\epsilon, e}/\kT)/Z(\beta)$, we get
\begin{equation}
\frac{p^\infty_{\beta_0}[\epsilon_\Sigma(t_0)]}{p^\infty_{\beta_\Sigma(t_N)}[\epsilon_\Sigma(t_N)]} = \exp((\Delta {\cal E}_\text{q}\Traj - \Delta F\Traj)/\kT).
\end{equation}
The initial and final thermal distributions respectively depend on $\beta_0$ and $\beta_\Sigma(t_N)$, which leads to a trajectory-dependent free energy variation $\Delta F\Traj = k_\text{B} T \log(Z(\beta_0)/Z(\beta_\Sigma(t_N)))$. 
Finally,
\begin{align}
\ds\Traj 
&= -\log(p_\text{m}[\beta_\Sigma(t_N)])  + \frac{\Delta {\cal E}_\text{q}\Traj - \Delta F\Traj - Q\Traj}{\kT} \nonumber\\
&= I_\text{Sh}\Traj -\frac{ \Delta \mathcal{E}_\text{m} + \Delta F\Traj}{\kT}\nonumber\\
&= I_\text{Sh}\Traj + \sigma\Traj,
\end{align}
where we used $\Delta {\cal E}_\text{q}\Traj = W\Traj + Q\Traj$ and $W\Traj = -\Delta \mathcal{E}_\text{m}\Traj$ [Eq.~(11) from the main text].

\supplsection{Reduced Jarzynski Equality}
We show that, in the Markovian limit, the reduced entropy production $\sigma [\rightvect{\Sigma}\,]$ obeys the Jarzynski like equality:
\begin{equation}
\ev{\exp(-\sigma[\rightvect{\Sigma}\,])} _{\rightvect{\Sigma}}= 1.  \label{suppl:JE}
\end{equation}
The derivation starts from the sum over all reversed trajectories of the complete machine: $1 = \sum_{\sleftvect{\Sigma}} \tilde{P}\rTraj$. In the limit $|\beta_0| \gg g_{\text{m}}/\Omega$, the action of the MO on the qubit is similar to an external operator imposing the evolution of the qubit frequency $\omega(\beta_0(t))$. As a consequence, the reversed jump probability at time $t_n$ does not depend on the exact MO state $\beta_\Sigma(t_n)$, but only on $\beta_{0}(t_n) = \beta_0\e^{-\ii\Omega t_{n}}$, which corresponds to the free MO dynamics. Therefore, we can get rid of the state dependencies in the MO state: $\tilde{P}[\Psi_\Sigma(t_{n-1}) | \Psi_\Sigma(t_{n}) ] =  \tilde{P}[\epsilon_\Sigma(t_{n}) | \epsilon_\Sigma(t_{n+1})]$ and $p^\infty_{\beta_\Sigma(t_N)}[\epsilon_\Sigma(t_N)] = p^\infty_{\beta_0(t_N)}[\epsilon_\Sigma(t_N)]$. Therefore,
\begin{align*}
1 
=\,& \left(\sum_{\beta_\Sigma(t_N)} p_\text{m}[\beta_\Sigma(t_N)]\!\right)\sum_{\rightvect{\epsilon}}p^\infty_{\beta_0(t_N)}[\epsilon_\Sigma(t_N)] \prod_{n = 1}^{N}  \tilde{P}[\epsilon_\Sigma(t_{n-1}) | \epsilon_\Sigma(t_{n})]\\
=\,&  \sum_{\rightvect{\epsilon}}p^\infty_{\beta_0(t_N)}[\epsilon_\Sigma(t_N)] \prod_{n = 1}^{N}  \tilde{P}[\epsilon_\Sigma(t_{n-1}) | \epsilon_\Sigma(t_{n})]\\
=\,& \sum_{\rightvect{\epsilon}} P[\rightvect{\epsilon}\,] \frac{p^\infty_{\beta_0(t_N)}[\epsilon_\Sigma(t_N)] \prod_{n = 1}^{N}   \tilde{P}[\epsilon_\Sigma(t_{n-1}) | \epsilon_\Sigma(t_{n})]}{p^\infty_{\beta_0}[\epsilon_\Sigma(t_0)] \prod_{n = 1}^{N}P[\epsilon_\Sigma(t_{n}) | \epsilon_\Sigma(t_{n-1})]}.
\end{align*}
Since the trajectory of the MO $\vec{\beta}[\rightvect{\epsilon}\,]$ is completely determined by the one of the qubit, we can restore the sum over the trajectories $\rightvect{\Sigma}$ of the autonomous machine. Then, from the expressions $p^\infty_{\beta}[\epsilon] = \exp(-\hbar\omega(\beta)\delta_{\epsilon, e}/\kT)/Z(\beta)$,  $W\Traj = -\Delta {\cal E}_\text{m}\Traj$ and $\tilde{P}[\Psi_\Sigma(t_{n-1}) |\Psi_\Sigma(t_{n})]/P[\Psi_\Sigma(t_{n}) |\Psi_\Sigma(t_{n-1})] = \exp(-q[\Sigma, t_{n - 1}])$ , we get
\begin{align*}
1 
&= \sum_{\rightvect{\Sigma}} P\Traj \exp(-\frac{\Delta {\cal E}_\text{q}\Traj - \Delta F - Q\Traj}{k_\text{B}T})\\
&=\ev{\exp(\frac{\Delta {\cal E}_\text{m}\Traj + \Delta F}{\kT})}_{\srightvect{\Sigma}}.
\end{align*}

\supplsection{Fluctuation theorem for the complete autonomous machine.}

\begin{figure}[!htb]
    \includegraphics[width=0.995\linewidth]{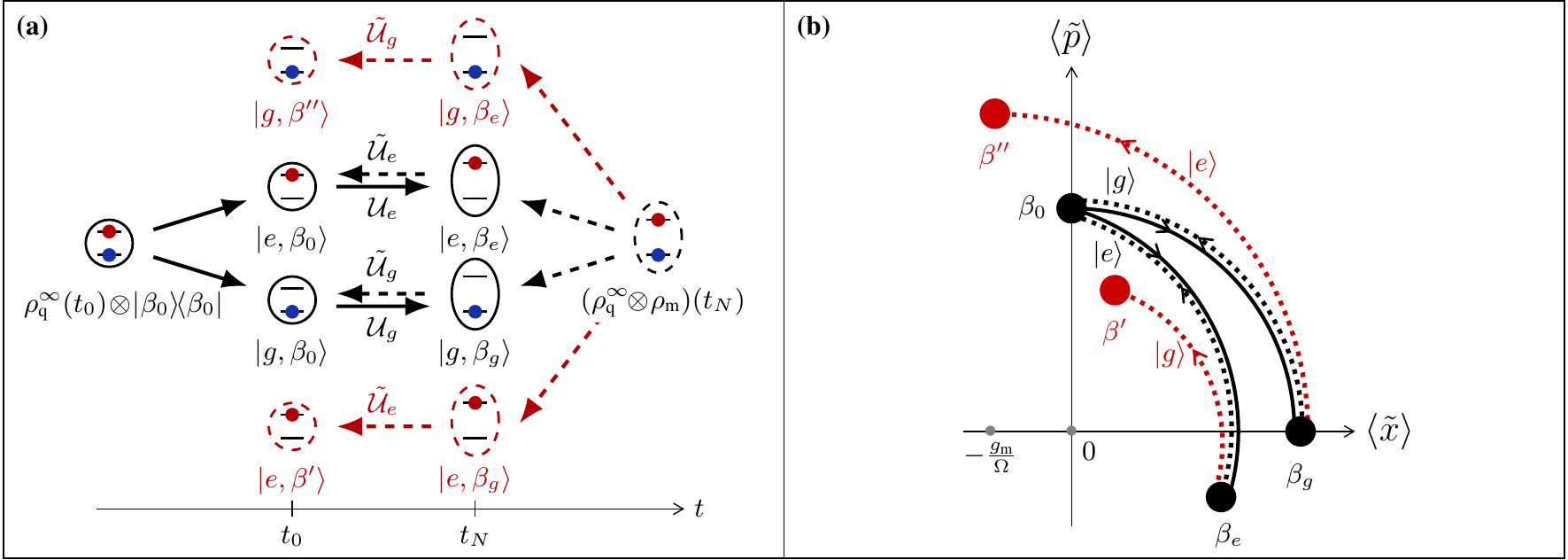}
    \caption{\label{fig-supp1} Example of trajectories for the qubit \textbf{(a)} and the MO \textbf{(b)}. The solid (resp. dashed) arrows correspond to the direct (resp. reversed) protocol. For the sake of simplicity, only the trajectories without any jump are represented. $\beta_g$ (resp. $\beta_e$) is the final state of the MO after the direct protocol when the qubit is in state $\ket{g}$ (resp. $\ket{e}$). The expressions of the MO evolution operators are: ${\cal U}_\epsilon(t) = \exp(-\ii t H_\text{m}^\epsilon)$ and $\tilde{\cal U}_\epsilon(t) = {\cal U}_\epsilon^\dagger(t)$, with $\epsilon=e, g$. The reversed trajectories that do not have a direct counterpart are plotted in red and the corresponding qubit states with dashed lines. The final MO states for these trajectories are  $\ket{\beta''} = \tilde{\cal U}_g(t_N)\ket{ \beta_e}$ and $\ket{\beta'} = \tilde{\cal U}_e(t_N)\ket{ \beta_g}$, where $\beta''\neq \beta_0$ and $ \beta' \neq \beta_0$. $\rho^\infty_\text{q}(t)$ (resp. $\rho_\text{m}(t)$) is the qubit thermal state (resp. the MO average state) at time $t$.}
\end{figure}

The IFT for the complete autonomous machine [Eq.~(23) from the main text] can be derived starting from the sum over all reversed trajectories, making appear the ratio $\tilde{P}\rTraj/P\Traj$. To do so, we need to ensure that $P\Traj \neq 0$. This requires to separate the set $\Sigma_\text{d} = \{\tilde{P}\rTraj | P\Traj \neq 0\}$ of reversed trajectories with a direct counterpart from the set without:
\begin{equation}
1 =\sum_{\sleftvect{\Sigma}} \tilde{P}\rTraj=\sum_{\sleftvect{\Sigma} \in \Sigma_\text{d}} P\Traj\frac{\tilde{P}\rTraj}{P\Traj} + \sum_{\sleftvect{\Sigma} \notin \Sigma_\text{d}} \tilde{P}\rTraj.
\end{equation}
Only the reversed trajectories $\leftvect{\Sigma} = \{ |\tilde{\epsilon}_\Sigma(t_n), \tilde{\beta}_\Sigma(t_n)\rangle \}_{n=N}^0$ such that $\tilde{\beta}_\Sigma(t_0) = \beta_0$ verify $P\Traj \neq 0$.
Fig.~\ref{fig-supp1} gives examples of both kinds of trajectories. Denoting $\lambda = \sum_{\sleftvect{\Sigma} \notin \Sigma_\text{d}} \tilde{P}\rTraj$ and using Eqs.~(13) and (17) from the main text we obtain:
\begin{align}
1 
=\,& \sum_{\rightvect{\Sigma}} \left(P\Traj p_\text{m}[\beta_\Sigma(t_N)]\frac{p^\infty_{\beta_\Sigma(t_N)}[\epsilon_\Sigma(t_N)]}{p^\infty_{\beta_0}[\epsilon_\Sigma(t_0)]}\frac{ \prod_{n = 1}^{N}\tilde{P}[\Psi_\Sigma(t_{n-1}) | \Psi_\Sigma(t_{n}) ]}{\prod_{n=1}^{N} P[\Psi_\Sigma(t_{n}) | \Psi_\Sigma(t_{n-1}) ]}\right) + \lambda\nonumber\\
=\,& \sum_{\srightvect{\Sigma}} P\Traj\exp(-I_\text{Sh}\Traj- \frac{\Delta {\cal E}_\text{q}\Traj- \Delta F\Traj - Q\Traj}{\kT}) + \lambda\nonumber\\
=\,&\ev{\exp(-(\sigma\Traj + I_\text{Sh}\Traj))}_{\srightvect{\Sigma}} + \lambda.
\end{align}
Thus, $\ev{\exp(-\Delta_\text{i} s\Traj)}_{\rightvect{\Sigma}} = 1- \lambda.$
\end{document}